\documentclass[apj, twocolumn]{aastex61}

\usepackage{url}
\usepackage{tabularx}
\usepackage{booktabs}
\usepackage{color}
\usepackage{amsmath}
\usepackage{hyperref}
\usepackage{upgreek}

\newcommand{\psr}{PSR\,J0837--2454}
\newcommand{\eg}{e.\,g.\ }

\begin{document}

\title{The Location of Young Pulsar \psr: Galactic Halo or Local Supernova Remnant?}

\author{Nihan Pol}
\affiliation{Department of Physics and Astronomy, West Virginia University, Morgantown, West Virginia 26506, USA}
\affiliation{Center for Gravitational Waves and Cosmology, West Virginia University, Chestnut Ridge Research Building, Morgantown, West Virginia 26505}
\affiliation{Department of Physics and Astronomy, Vanderbilt University, 2301 Vanderbilt Place, Nashville, TN 37235, USA}
\author{Sarah Burke-Spolaor}
\affiliation{Department of Physics and Astronomy, West Virginia University, Morgantown, West Virginia 26506, USA}
\affiliation{Center for Gravitational Waves and Cosmology, West Virginia University, Chestnut Ridge Research Building, Morgantown, West Virginia 26505}
\affiliation{CIFAR Azrieli Global Scholar; Canadian Institute for Advanced Research, MaRS Centre West Tower, 661 University Ave. Suite 505, Toronto ON M5G 1M1, Canada}
\author{Natasha Hurley-Walker}
\affiliation{International Centre for Radio Astronomy Research, Curtin University, Bentley, WA 6102, Australia}
\author{Harsha Blumer}
\affiliation{Department of Physics and Astronomy, West Virginia University, Morgantown, West Virginia 26506, USA}
\affiliation{Center for Gravitational Waves and Cosmology, West Virginia University, Chestnut Ridge Research Building, Morgantown, West Virginia 26505}
\author{Simon Johnston}
\affiliation{CSIRO Astronomy and Space Science, Australia Telescope National Facility,
PO Box 76, Epping, NSW 1710, Australia}
\author{Michael Keith}
\affiliation{Jodrell Bank Centre for Astrophysics, University of Manchester, Alan Turing Building, Oxford Road, Manchester M13 9PL, United Kingdom}
\author{Evan F. Keane}
\affiliation{Centre for Astronomy, School of Physics, National University of Ireland Galway, University Road, Galway, H91 TK33, Ireland.
}
\author{Marta Burgay}
\affiliation{INAF - Osservatorio Astronomico di Cagliari, via della Scienza 5, 09047 Selargius (CA), Italy}
\author{Andrea Possenti}
\affiliation{INAF - Osservatorio Astronomico di Cagliari, via della Scienza 5, 09047 Selargius (CA), Italy}
\affiliation{University of Cagliari, Dept of Physics, S.P. Monserrato-Sestu Km 0,700 - 09042 Monserrato (CA), Italy}
\author{Emily Petroff}
\affiliation{Anton Pannekoek Institute for Astronomy, University of Amsterdam, Science Park 904, 1098 XH Amsterdam, The Netherlands}
\author{N.D. Ramesh Bhat}
\affiliation{International Centre for Radio Astronomy Research, Curtin University, Bentley, WA 6102, Australia}

\begin{abstract}
    We present the discovery and timing of the young (age $\sim 28.6$~kyr) pulsar \psr. Based on its high latitude ($b = 9.8^{\circ}$) and dispersion measure (DM $ = 143$~pc~cm$^{-3}$), the pulsar appears to be at a $z$-height of $>$1\,kpc above the Galactic plane, but near the edge of our Galaxy. This is many times the observed scale height of the canonical pulsar population, which suggests this pulsar may have been born far out of the plane. 
    If accurate, the young age and high $z$-height imply that this is the first pulsar known to be born from a runaway O/B star.
    In follow-up imaging with the Australia Telescope Compact Array (ATCA), we detect the pulsar with a flux density $S_{1400} = 0.18 \pm 0.05$~mJy. We do not detect an obvious supernova remnant around the pulsar in our ATCA data, but we detect a co-located, low-surface-brightness region of $\sim$1.5$^\circ$ extent in archival Galactic and Extragalactic All-sky MWA Survey data. 
    We also detect co-located H$\alpha$ emission from the Southern H$\alpha$ Sky Survey Atlas. Distance estimates based on these two detections come out to $\sim$0.9~kpc and $\sim$0.2~kpc respectively, both of which are much smaller than the distance predicted by the NE2001 model \citep[$6.3$~kpc,][]{ne2001} and YMW model \citep[$>25$~kpc,][]{ymw17} and place the pulsar much closer to the plane of the Galaxy.
    If the pulsar/remnant association holds, this result also highlights the inherent difficulty in the classification of transients as ``Galactic'' (pulsar) or ``extragalactic'' (fast radio burst) toward the Galactic anti-center based solely on the modelled Galactic electron contribution to a detection. 
\end{abstract}


\section{Introduction}
    
    Pulsars form in supernovae (SNe), resulting from the core collapse of massive stars. This explosion creates a supernova remnant (SNR), which, if not too distant, can be detectable for thousands of years. This evolutionary association of pulsars in supernova births leads to two clear observational effects: first, both SNR and young pulsars are observed at low Galactic latitudes $\lesssim300\,$pc, where the bulk of O/B-type stars reside \citep{OB_star_spatial_dist_1}; and second, we find that many young pulsars are associated with an SNR; this appears to be the case for pulsars with ages of up to a few tens of kyr \citep[][]{snr_psr_1, snr_psr_2, snr_psr_4, snr_psr_5, 1643-43_snr_psr}.
    
    SNRs are excellent tools for probing various properties of the interstellar medium and the pulsar itself. The presence of an SNR can provide information about the local electron density \citep[e.g.,][]{frail_snr_ism_prop} and magnetic field strength of the pulsar \citep[e.g.,][]{snr_spin_p_B_fields}. The size of the SNR along with measurement of its expansion velocity can provide an estimate of the age of the pulsar. The relative position of the pulsar with respect to the SNR can be used to infer the velocity of the pulsar, and thus, provide estimates on the kick received by the pulsar at the time of the SN explosion \citep{snr_proper_velocity_use}.
    
    In this publication, we report the discovery, timing, and follow-up of a pulsar that appears, based on the distance derived from its dispersion measure (DM), to lie at least 1\,kpc above the outer arms of the Galaxy.
    Based on the current period, $P$, and period-derivative, $\dot{P}$, of the pulsar, this pulsar appears to be young, with its characteristic age given by \citep[][]{psr_handbook}
    \begin{equation}
        \displaystyle \tau_{\rm c} = \frac{P}{(n - 1)\dot{P}} \left[ 1 - \left( \frac{P_{\rm birth}}{P} \right)^{n - 1}  \right]
        \label{char_age}
    \end{equation}
    where $n$ is the braking index of the pulsar and $P_{\rm birth}$ is the spin period of the pulsar when it was born. Given the large period of \psr\ ($P = 629.4102$~ms), we can safely assume that $P_{\rm birth} \ll P$. Assuming that the pulsar spin-down is from magnetic-dipole braking only, the braking index is $n = 3$, and Eq.~\ref{char_age} reduces to 
    \begin{equation}
        \displaystyle \tau_{\rm c} = \frac{P}{2 \dot{P}} \approx 28.6 \, {\rm kyr}
        \label{approx_char_age}
    \end{equation}
    The young age and high Galactic latitude of this pulsar are contrary to the standard formation scenario described above. The location of this pulsar can be explained via two different scenarios: (i) it was born at this high Galactic latitude by a runaway O/B star; or (ii) it was born closer to the plane of the Galaxy and the distance derived using the NE2001 model \citep{ne2001} is incorrect.
    
    The presence of a supernova remnant around the pulsar would allow us to independently estimate the distance to the pulsar-SNR system using the Sigma-D relation \citep{sigma-D_relation}. In addition, there are only six confirmed Galactic supernovae that are above a Galactic latitude of $b \gtrsim 8.5$ \citep{green_snr_cat}, while only two of these have a neutron star at their center \citep{ns_in_highlat_snr_2, ns_in_highlat_snr_1}. However, both of these neutron stars are radio-silent, and thus, a supernova remnant detection around this pulsar would be a valuable addition to this sample.
    Assuming the characteristic age is a reasonable assumption of the true age of the pulsar and SNR, we would expect the SNR to be in its late Sedov \citep{sedov_expansion} or early radiative evolution phase \citep{radiative_phase_snr} and thus would see synchrotron emission at radio wavelengths from the SNR, with the possibility of further emission at X-ray wavelengths \citep{handbook_sne}. 
    
    If the pulsar is indeed as high as the DM-derived distance, that would imply that the progenitor of this pulsar was likely a runaway O/B star. Runaway stars \citep{first_runaway_star} are stars that were originally part of a binary which was disrupted either when the other star in the binary explodes as a supernova \citep[the Hills mechanism,][]{Hills_mechanism} or due to dynamical three- or four-body interactions \citep[][]{three_body_scatt_1, three_body_scatt_2}. The disruption of the binary system results in runaway stars having peculiar velocities greater than 30~km~s$^{-1}$ \citep{runaway_star_numerical_sim}.
    Since most of the O/B stars are located close to the Galactic plane \citep{OB_star_spatial_dist_1} and given the young characteristic age of the pulsar, it would have to be born close to its current position. Thus, the progenitor of the pulsar would have to be a runaway star to be able to reach its current observed position. If we can confirm that the position of the pulsar as predicted by the NE2001 model is accurate, this would be the first known pulsar whose progenitor was a runaway star.
    
    This paper is structured as follows. In Section~\ref{obs_and_results}, we describe the discovery and timing of \psr, and our observations and analysis of the data collected with ATCA. In Section~\ref{results}, we present the results from the timing and interferometric analysis. In Section~\ref{discuss}, we discuss the implications of our results and present the conclusion in Section~\ref{conclusion}.

\section{observations and data} \label{obs_and_results}

\subsection{Discovery and Timing of \psr}\label{sec:timing}

\begin{figure}
\centering
\includegraphics[trim=30mm 23mm 13mm 23mm,clip,width=0.48\textwidth]{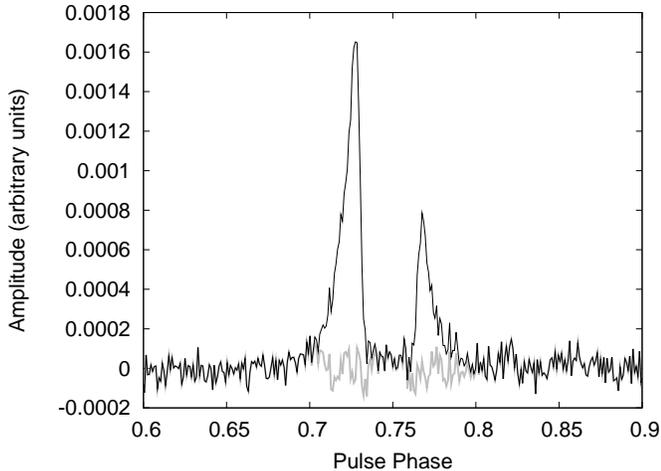}
\caption{In black we show the integrated profile of \psr\ at reference frequency of 1382~MHz for the BPSR/HIPSR data over the period January, 2012 to September, 2015.
The grey line shows the result of subtracting our analytic timing template from this integrated profile. For clarity of this figure, the analytic template is not shown.}
\label{fig:2012-2015}
\end{figure}

\psr\ was discovered as part of the High Time Resolution Universe (HTRU) intermediate-latitude survey \citep[HTRU; ][]{HTRU} carried out with the Parkes Radio Telescope in both the single-pulse and periodicity pipelines \citep{HTRU-III}. No period has previously been reported for this pulsar. Its timing solution, derived as described below, is shown in Table\,\ref{table:timingsolution}.

After its discovery, temporally resolved spectra (``search data'') were obtained for this pulsar from the time of its discovery in 2011, until early 2017. For this period of time, the pulsar was not observed using standard timing  observations (``folded data''), because of the difficulty in obtaining a coherent timing solution. This was due largely to its large red noise properties, and in part to its inaccurate position and large period derivative (which initially quickly made the pulsar lose phase coherence). All search data were collected at frequencies within the 1.1--1.5\,GHz band using the BPSR \citep[Berkeley Parkes Swinburne Recorder,][]{BPSR}, HIPSR \citep[HI-Pulsar,][]{HIPSR}, and DFB4 \citep[Digital Filterbank 4][]{DFB4} instruments; this data is available for download from CSIRO's online Data Access Portal\footnote{https://data.csiro.au/dap/}. After flagging of interference-affected channels, the average frequency of most data, and the reference frequency we used for timing, was 1382\,MHz.

We measured an initial timing solution using single-pulse events from all epochs. It was then noted that the pulsar was also visible in a Fourier search; to obtain TOAs with higher stability (i.e. not subject to the pulse-to-pulse variations that are well-known to occur in pulsars), each observation was then averaged over the rotational period of the pulsar (``folded'') using that single-pulse solution. A time-of-arrival (TOA) was measured separately for each observation, using the {\sc dspsr} \citep{dspsr} and {\sc psrchive} \citep{psrchive} softwares. TOA creation requires comparing new observations to a standard template.
We first tried timing all data against a rudimentary template created from a subset of the data, summing all detections over the time period from June to September 2011. A template used in this way is not ideal, as the cross-correlation of noisy data with a template made with that same data results in erroneously low TOA errors. Thus, we used an analytic template, which we obtained as described below.

We first aimed to produce a higher-S/N integrated profile with which to fit a representative analytic template. As with many young pulsars, \psr\ is dominated by large amounts of red noise \citep[\eg][]{timing-noise}. Thus, we were only able to obtain a low root-mean-squared (RMS) timing solution (RMS residuals 613\,$\upmu$s) over a sub-set of data in the time frame January, 2012 to September, 2015, fitting for RA, Dec, $F$, $\dot{F}$, and $\ddot{F}$ (which correspond to right ascension, declination, pulsar spin frequency, frequency derivative, and second frequency derivative, respectively). We integrated the folded data across the BPSR/HIPSR data during these epochs to obtain a high-S/N profile. These data, integrated over that entire time span, produce the profile seen in Figure~\ref{fig:2012-2015}.
To that profile, we fit an analytic template and re-derived TOAs and a timing solution for all data. The analytic template consisted of four Gaussian components, and its subtraction from the data left the noise-like residual shown in grey in Fig.~\ref{fig:2012-2015}.

\begin{figure}
\centering
\includegraphics[width=0.48\textwidth]{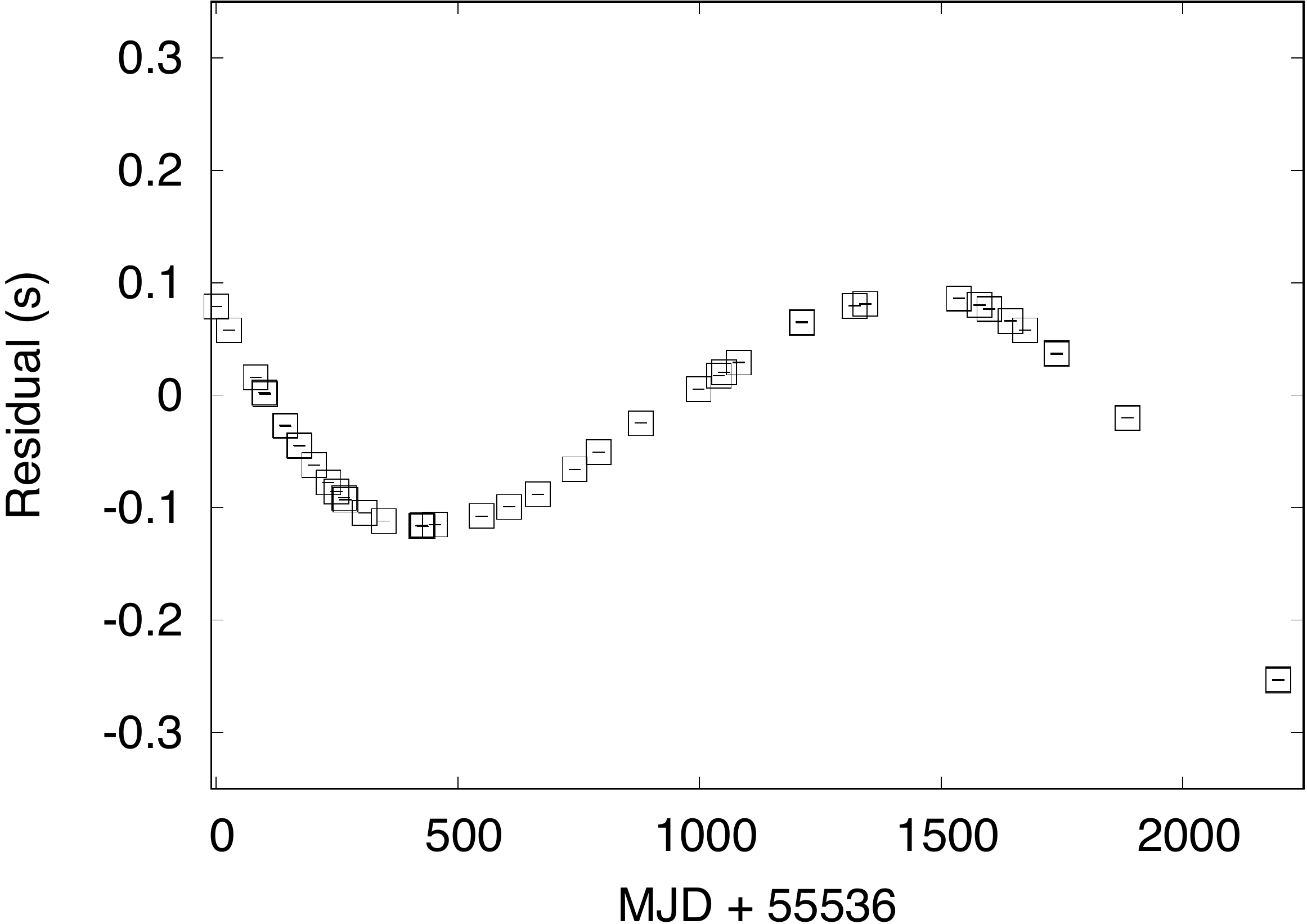}
\caption{Here we show the residuals following the timing procedure described in Section \ref{sec:timing}. A clear cubic term, encompassing part of the red noise exhibited by this pulsar, is clear. The error bars are plotted but are sufficiently small ($<10\,\upmu$s) such that they are not clearly visible on this plot.}
\label{fig:timingsolution}
\end{figure}

Due to the significant red noise observed in this pulsar (large power in low-order terms; see Fig.\,\ref{fig:timingsolution}), we employed the method of \citet{coles+11} to solve for the pulsar parameters using a generalized least-squares approach. We used the {\sc SpectralModel} plugin of the {\sc tempo2} timing software to model the red noise as a power-law with spectral index of $-6$, typical of the steep red noise process commonly seen in young pulsars.

In the residuals that result from the final timing solution, there is clearly a large cubic term apparent in the data which causes the pulsar's rotation to vary greatly during our several-year data span. The root-mean-squared residual variance in our final timing solution is 90\,ms, 
which is dominated by the red noise (individual TOA error bars are on the order of 10\,$\upmu$s, and as previously noted a timing solution over a sub-set of the full time span resulted in a root-mean-squared residual variance of a more reasonable 613\,$\upmu$s).

The observed and derived parameters from our observations and from our timing solution are presented in Table \ref{table:timingsolution}.

\begin{table}\label{table:timingsolution}\caption{Timing solution  for \psr.}
\centering
\begin{tabular}{ll}
\hline
{\bf Parameter} & {\bf Value (err)} \\ 
\hline
\\
{\bf Measured Parameters}\\
Right ascension (J2000) & 08:37:57.73(6)\\
Declination (J2000) & -24:54:30(1)\\
Galactic longitude ($^\circ$) & 247.58 	\\
Galactic latitude ($^\circ$) & 9.77\\
DM (pc\,cm$^{-3}$) & 143.1(1)\\
$P$ (ms)& 629.41024(2)\\
$\dot P$ ($\times10^{-13}$ s/s) & 3.490(2)\\    
$F$ (1/s) & 1.58878890(5)\\
$\dot F \ \rm (1/s^2)$& $-8.808(4)\times10^{-13}$\\
\\
{\bf Derived Parameters}\\
$\tau_{\rm c}$ (kyr) & 28.6\\
$B_{\rm surf}$ (G) & $1.5\times 10^{13}$\\
$\dot E$ (erg\,s$^{-1}$) &  $5.5\times10^{34}$\\
DM distance (NE2001) & 6.29\,kpc\\
$z$-height (NE2001) & 1.1\,kpc\\
DM distance (YMW17) & $>$25\,kpc\\
$z$-height (YMW17) & ---\\
\\
{\bf Reference Values}\\
Reference frequency (MHz)  & 1381.999\\
Reference epoch (PEPOCH, MJD)  & 55588.628331\\
TOA range (MJD) & 55536-57787\\
Solar System Ephemeris & DE405\\
Units & TCB (tempo2)\\
\\
\hline
\end{tabular}\\
\end{table}

\subsection{ATCA Observations and Data Reduction}
    
    We used the Australia Telescope Compact Array (ATCA) in the 1-3 GHz band to observe the field of \psr\ and search for any supernova remnant (SNR) that might be associated with the pulsar. We were awarded 24 hours, split into two observing sessions of 12 hours each. The first of these sessions was conducted on November 4, 2011 with the array in 750C configuration. The second was on March 4, 2012 with the array in EW352 configuration. These observations provided a balance between a large field of view and ideal frequency placement to detect what would be a non-thermal power-law spectrum of the SNR.  The observations were centered on an early solution for the position of the pulsar, RA 08:37:57.76 and DEC -24:54:30.51 (J2000 epoch), which has a small offset of 0.08~s in RA from the final timing solution reported in Table~1.
    
    We used J1934--638 as an ATCA standard primary calibrator and J0843--259 as the phase calibrator for both the observing sessions. A total of $\sim$1.5 hours across the two sessions were spent observing the primary calibrator. We spent a net sum of $\sim$4 hours observing the phase calibrator; those observations were interleaved with on-source observations. This leaves a total of 18.5 hours on-target. Both the sessions were carried out with a center frequency of 2.1~GHz, and 2048 frequency channels, with each channel 1 MHz in width. This setup allows us to detect structures in the combined data from a smallest angular resolution of $\sim 5.8''$ to spatial scales as large as $\sim 15.8'$, with a field-of-view of $\sim 22.3'$.
    
    \begin{figure*}
        \centering
        \includegraphics[height =\textwidth,angle=270,trim=8mm 0mm 0mm 0mm,clip]{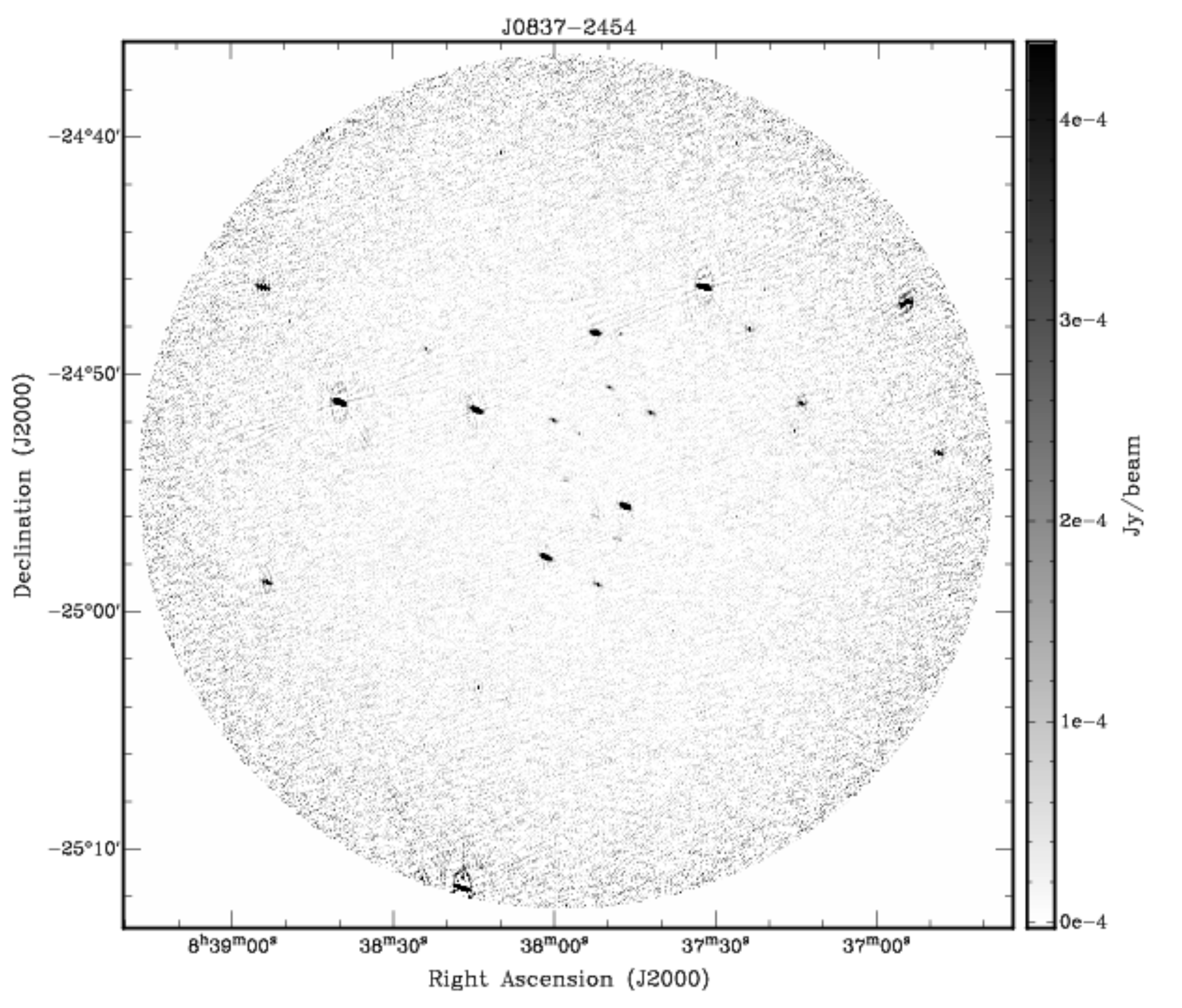}
        \caption{Image produced using uniform weighting in the deconvolution process. This image provides the highest spatial resolution but has a reduced average signal to noise ratio. This type of weighting is ideally suited for identifying point sources in the field of view. The pulsar \psr\ is weakly visible at the center of the image.}
        \label{uniform_wt_image}
    \end{figure*}
    
    \begin{figure*}
       \centering
        \includegraphics[height =\textwidth,angle=270,trim=4mm 0mm 0mm 0mm,clip]{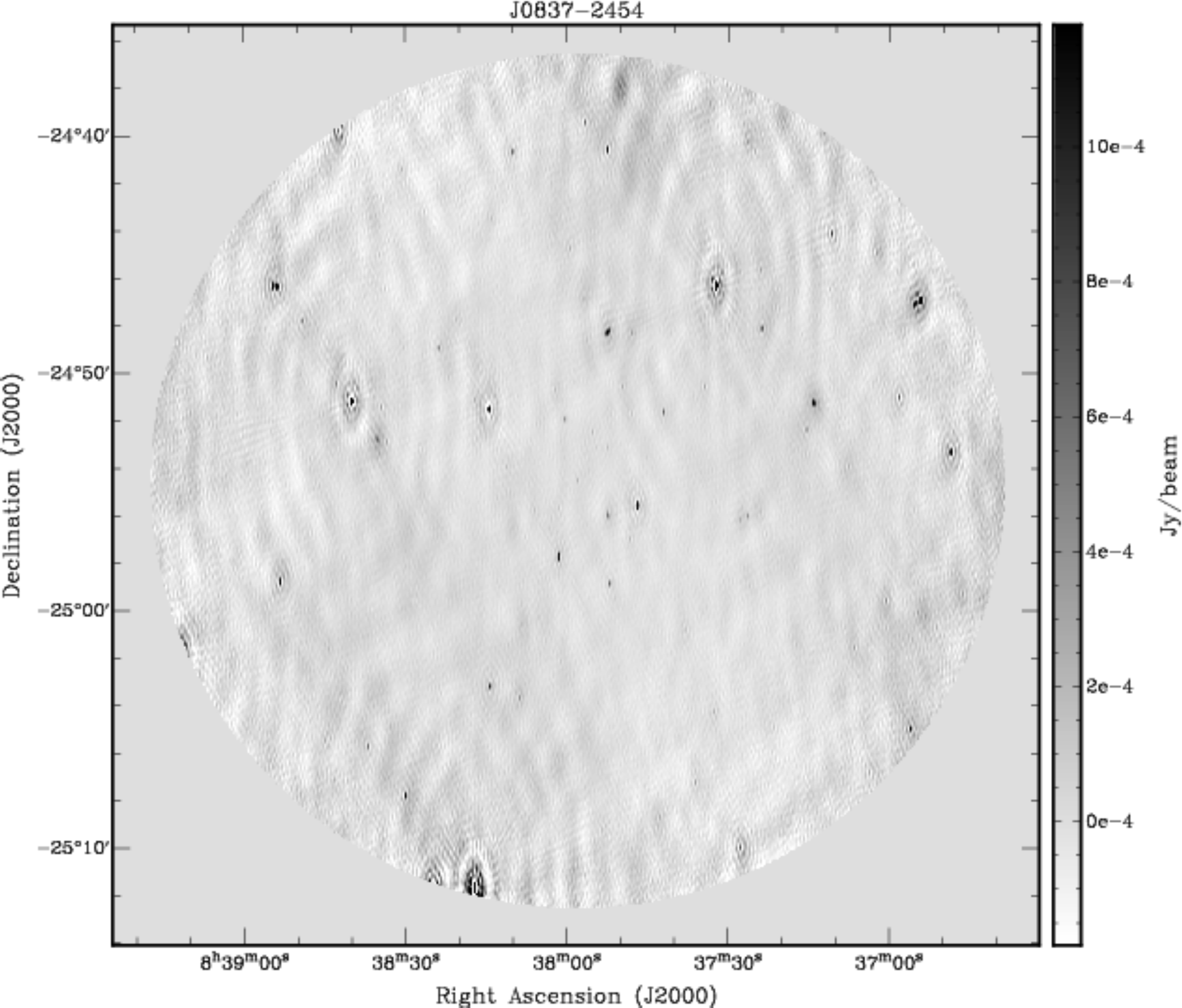}
        \caption{Image produced using Briggs weighting with robust parameter $R = 0$ in the deconvolution process. This image provides a good trade-off between spatial resolution and sensitivity, allowing us to search for a supernova remnant over a large range of spatial resolution. We do not detect any supernova remnant associated with \psr\ in this image.}
        \label{briggs_wt_image}
    \end{figure*}
    
    The data calibration and image processing was done using the CASA software package. The data from each epoch were manually flagged before being manually calibrated using standard CASA procedures\footnote{The procedure can be found at this URL: \url{https://casaguides.nrao.edu/index.php/CASA\_Guides:ATCA\_Advanced\_Continuum\_Polarization\_Tutorial\_NGC612-CASA4.7}}. The calibrated data from both observing sessions were concatenated into a single dataset. This concatenated dataset was then deconvolved and cleaned using CASA's {\sc clean} routine.
    
    We deconvolved the data using two different weighting techniques corresponding to the type of information we required from the image. First, we deconvolved the image using the uniform weighting  scheme implemented in CASA. This type of weighting returns an image of the sky with the best resolution, but with a reduced sensitivity. The image produced with uniform weighting is shown in Fig.~\ref{uniform_wt_image}. This image has a RMS flux of $\sigma = 24 \, \upmu$Jy.
    
    We also produced an image using the Briggs weighting \citep[][]{Briggs_wt} scheme in CASA, with the robustness parameter $R = 0$. This weighting gives a good trade-off between resolution and sensitivity, which allows us to search for a supernova remnant over a suitably large range of angular resolutions. This image is shown in Fig.~\ref{briggs_wt_image}. This image has an RMS flux of $\sigma = 32 \ \upmu$Jy. Our sensitivities improve on the archival NVSS imaging of this field \citep{NVSS} by a factor of more than 14.
    
\begin{figure}
    \centering
    \includegraphics[trim=3.0cm 2cm 1.9cm 2.2cm,clip, width=0.99\columnwidth]{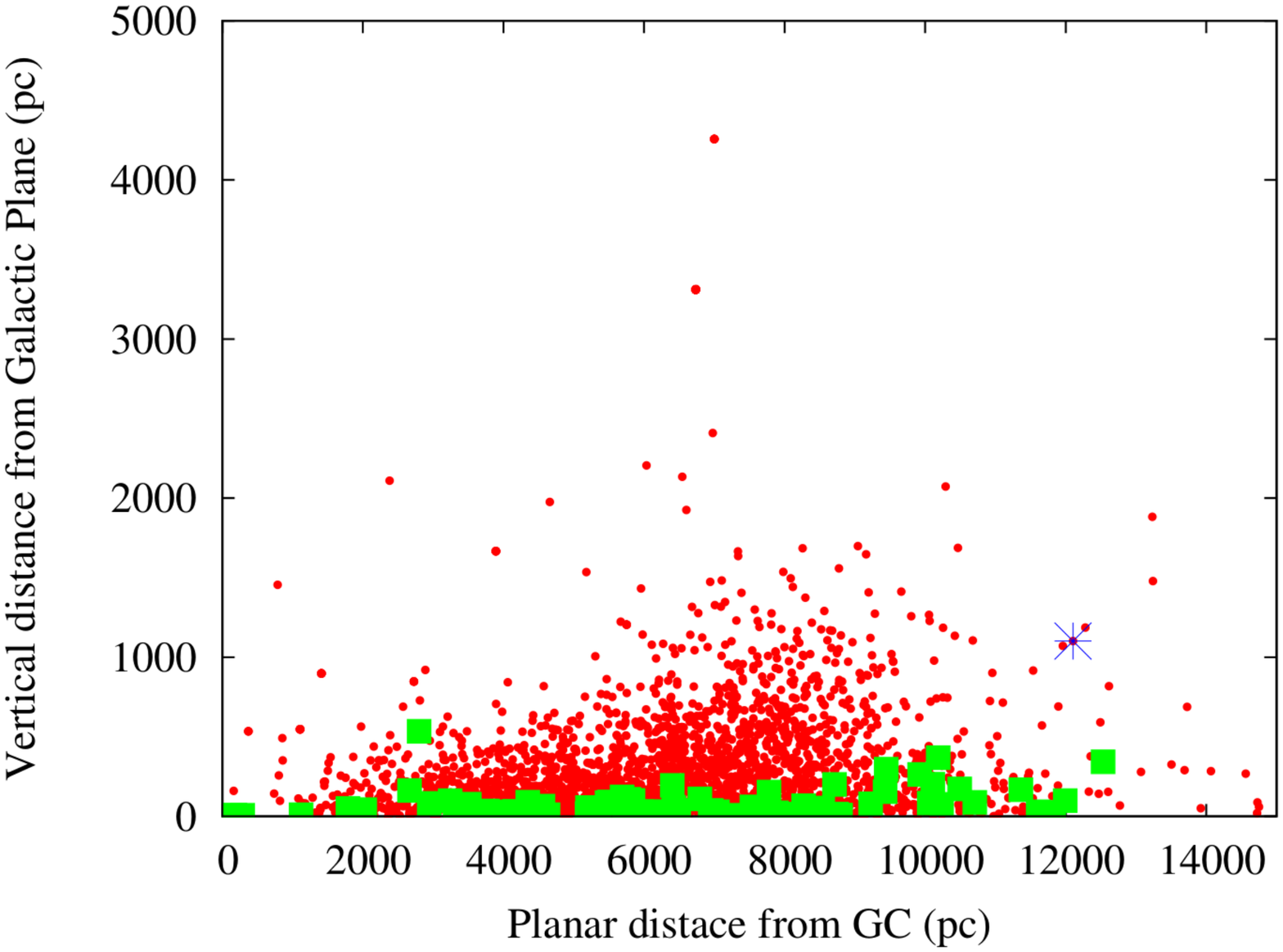}
    \caption{Here we show the positions of pulsars in the Galaxy relative to the Galactic Center (GC). The vertical axis represents the absolute value of the pulsar's $z$-height, while the horizontal axis represents the planar distance from the GC. For this plot we have assumed a GC-Sun distance of 8122\,pc, and drawn the galactic longitude and latitude ($gl, gb$ respectively) and NE2001-derived dispersion measure distance from {\sc psrcat}. All pulsars are shown as small red circles, pulsars of characteristic age less than 50\,kyr are shown as larger green squares, and \psr\ is shown as a blue asterisk.}
    \label{fig:gcdist}
\end{figure}

\section{Results} \label{results}
\subsection{Properties of the pulsar}
This pulsar stands out in several ways. First, its emission is highly modulated, with single pulses visible at a level of several hundreds of mJy, while the mean profile flux density is $\sim$80\,$\upmu$Jy.
We do not have sufficient S/N in our observations to assess whether there are any extended periods of nulling in the pulsar.

Using our highest-S/N observation, we measured the DM to be $143.1\pm0.1\,{\rm pc\,cm^{-3}}$. We also measured DM as a function of time (separately for each independent observation) and found no significant deviations within our measurement errors.

The pulsar is relatively young, and is at around the 13th percentile for young pulsar ages compared to cataloged pulsars with measured characteristic ages. The most striking thing about this pulsar is its large apparent height above the Galactic plane based on the NE2001-implied distance compared to the age of the pulsar. 
Given our measured DM, we can estimate the distance to the pulsar, $D$, based on the known Galactic free electron density, $n_e$, along the line of sight. Currently, the two most broadly used models of the Galactic free electron density, $n_e$, are the NE2001 model \citep[][]{ne2001} and YMW \citep[][]{ymw17} model. The distance prediction by the NE2001 model results in a radial distance $D_{\rm cl} = 6.3$~kpc, implying a height of this pulsar above the Galactic plane of 1.1\,kpc. The prediction from the YMW model indicates that there are only 118\,${\rm pc\,cm^{-3}}$ of Galactic DM in this direction, and thus puts the pulsar at the edge of the Galaxy (or, extragalactic, which is not likely, providing a prediction that the pulsar comes from a distance of $D_{\rm ymw} = 151$~Mpc). In Figure \ref{fig:gcdist}, we show the radial and orthogonal offset from the Galactic center for this pulsar, young pulsars ($\tau_{\rm c} < 50$\,kyr) and all pulsars as inferred from the DM-derived distances based on the {\sc NE2001} electron density model. While it is clear that this pulsar inhabits a somewhat sparsely populated position in the Galactic plane (specifically it is located well above the outer/Perseus arm of the Galaxy), the position in reference to the Galactic Center is even more pronounced when considering the young pulsar population that is comparable in characteristic age to this pulsar.
   
Most pulsars are born in supernova explosions in the disk of the Galaxy, and the explosive birth of a pulsar can impart a kick to the pulsar at birth, resulting in a high transverse velocity. Due to this kick, pulsars gradually migrate out of the plane of the Galaxy and populate the high latitude regions (i.e. the halo) of the Galaxy. The presence of a young pulsar, \psr, in the outer regions of the plane, as visible in Fig.\,\ref{fig:gcdist} is therefore surprising. 

\subsection{Objects in the ATCA field of view}
     
    We detect a point source at the position of \psr\ in both observing sessions. We also detect it in the concatenated data, with a flux of $0.18 \pm 0.05$~mJy ($5.4 \sigma$ significance). This flux density is consistent with that of the mean (continuum, non-pulsed) pulsar emission measured by Parkes telescope, if we use a duty cycle of $\sim$5\% that encompasses only the two emitting peaks of the pulsar's profile.
    
    We do not detect any extended emission in our ATCA data down to a level of 3 times the RMS image noise (equivalent to 96~$\upmu$Jy/beam) that might represent the existence of a compact SNR or pulsar wind nebula (PWN) around the pulsar. This non-detection places a conservative upper limit on the flux density of any non-pointlike features associated with the potential low-surface-brightness SNR of the GLEAM image (see Sec.~\ref{gleam_analysis}).
        
        We identified all the point sources in the uniform weighted image, associating them with any known sources in NED\footnote{\url{https://ned.ipac.caltech.edu/}} and SIMBAD \citep[][]{SIMBAD}. To identify objects in our field, we searched for point sources in the image with flux densities above $5\sigma$, where $\sigma = 24.43 \ \upmu$Jy is the RMS flux in the image. All of these sources, both with and without known associations in the NED and SIMBAD catalogs, are listed in the supplemental information provided with this paper. 
    
    \subsection{A search of archival data} \label{archival_data}
        
        \subsubsection{GLEAM survey} \label{gleam_analysis}
            
            \begin{figure}
                \centering
                \includegraphics[width=\columnwidth]{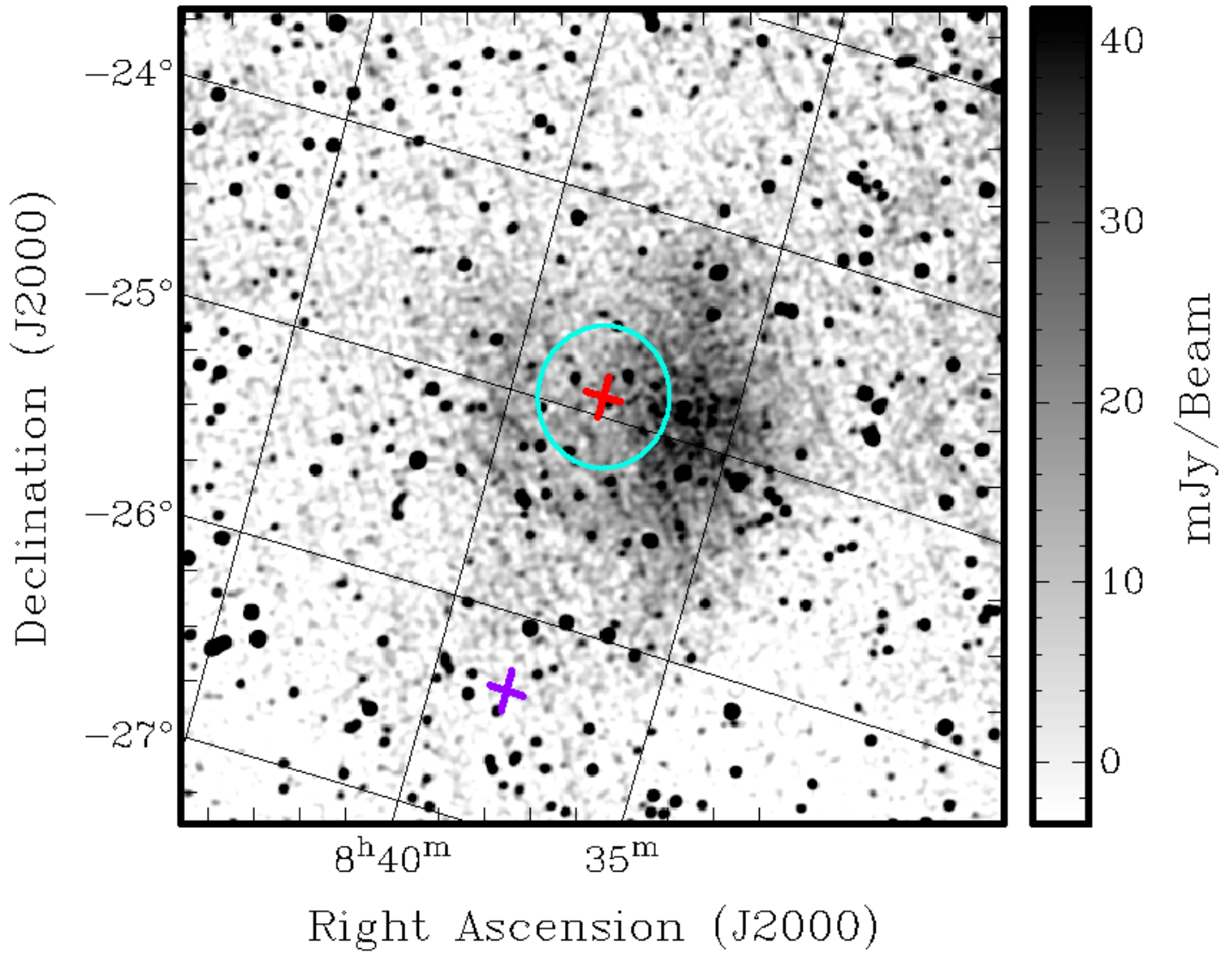}
                \caption{A wide-field view of the pulsar's position in the stacked 170--231\,MHz GLEAM data. The cyan circle shows the size of the image in Figures \ref{uniform_wt_image} and \ref{briggs_wt_image}. The crosses are of arbitrary size, and show the positions of pulsars \psr\ (red) and PSR J0838--2621 (purple).}
                \label{fig:gleam}
            \end{figure}
            
            We obtained images from the Galactic and Extragalactic All-sky MWA Survey (GLEAM; \citealt{GLEAM}) for this sky position\footnote{\url{http://gleam-vo.icrar.org/gleam_postage/q/form}}. Figure \ref{fig:gleam} shows a wide-field view of the stacked 170--231\,MHz images in this region, which exhibits a highly diffuse, semi-symmetric structure surrounding the pulsar position.
            
            Since the structures in the image have not been deconvolved due to a lack of multi-scale {\sc clean} when the images were generated, the choice of a background is essential in performing a peak brightness measurement across frequency to measure a spectral index for this structure. However, this is severely hampered due to the poorly defined edges of the source, as well as strong contamination from the large number of extragalactic sources in the image. As a result, it is hard to estimate the spectral index of the source with any amount of confidence. Consequently, the GLEAM data does not provide conclusive evidence of the diffuse structure being a SNR associated with \psr.
        
        \subsubsection{Southern H$\alpha$ Sky Survey (SHASSA)}
            \label{shassa_sec}
            \begin{figure*}
                \centering
                \includegraphics[scale = 0.5]{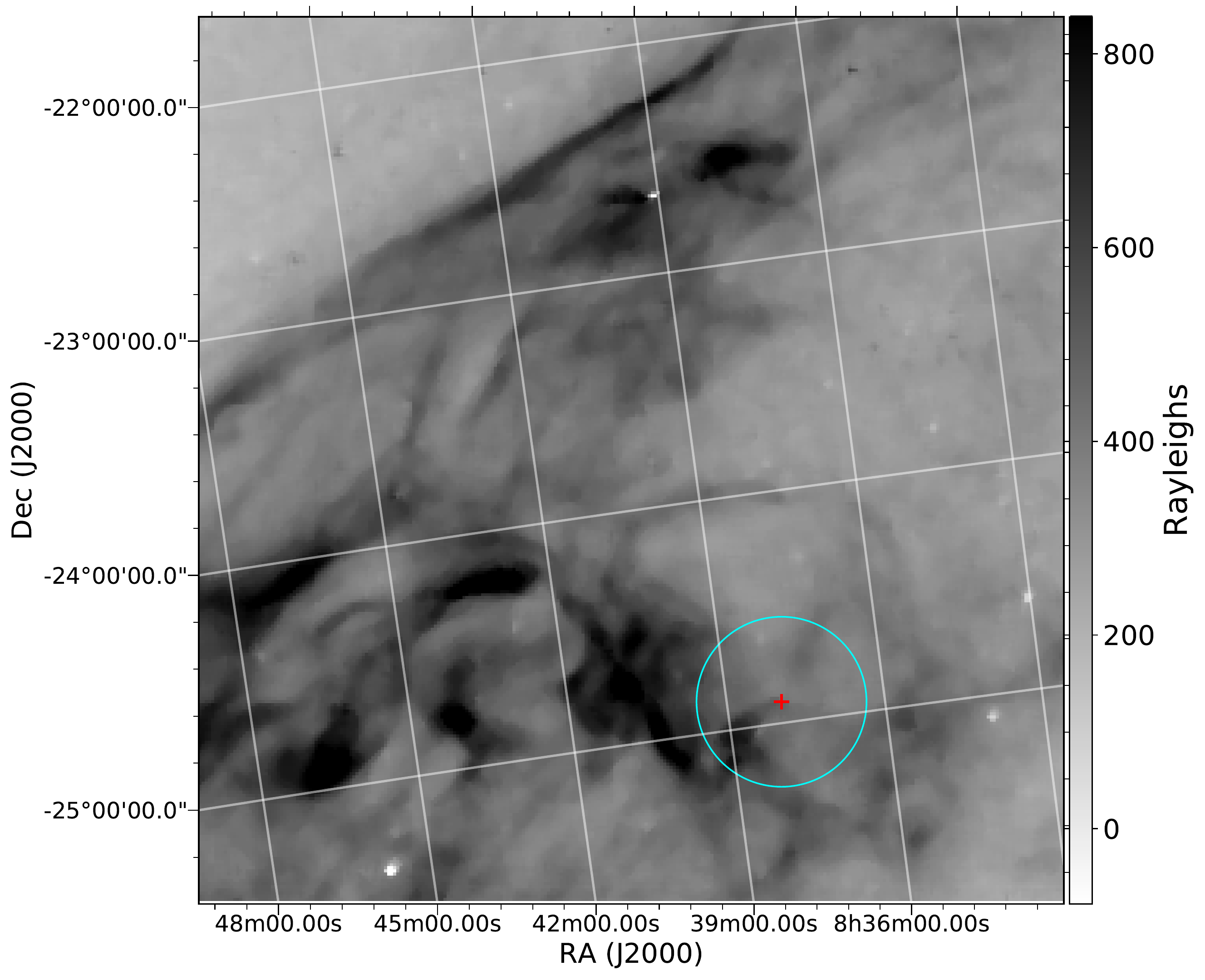}
                \caption{A wide-field view of the position of \psr\ obtained from the Southern H$\alpha$ Sky Survey (SHASSA). The position of the pulsar is shown by the red cross, while the cyan circle shows the size of the image in Figures \ref{uniform_wt_image} and \ref{briggs_wt_image}.}
                \label{shassa}
            \end{figure*}
            
            We obtained images from the Southern H$\alpha$ Sky Survey \citep[SHASSA,][]{shassa} for this sky position to search for associated H$\alpha$ emission from this region. The SHASSA image for this portion of the sky is shown in Figure~\ref{shassa}, with the position of \psr\ shown by the red cross and the cyan circle shows the size of the images in Figures \ref{uniform_wt_image} and \ref{briggs_wt_image}. As we can see, the location of \psr\ is coincident on the sky with a diffuse H$\alpha$ filament, with a photon flux amplitude of $\mathcal{R} = 470$~Rayleigh in the pixel at the position of the pulsar.
            The morphology of the surrounding H$\alpha$ emission is inconsistent with the GLEAM diffusion emission morphology and thus, we cannot conclusively claim that the two are related to each other.
            
        \subsubsection{Other archival data}
\begin{figure}
            \centering
            \includegraphics[trim=10mm 58mm 16mm 60mm,clip,width = \columnwidth]{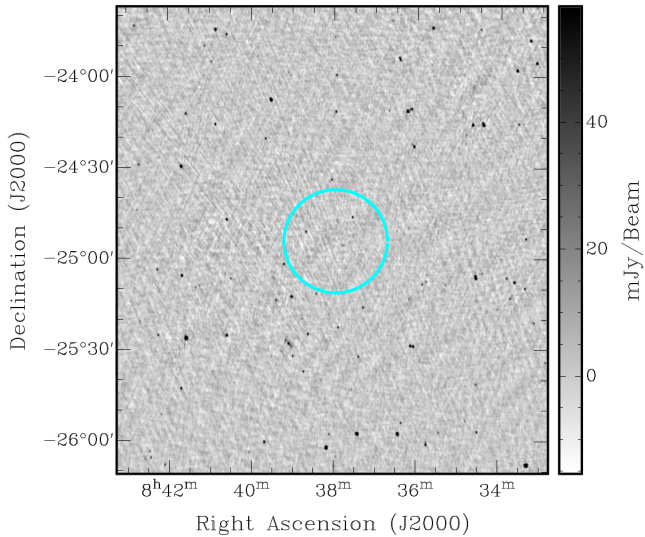}
            \caption{The 150\,MHz TGSS image in the region of \psr. The cyan circle shows the size of the image in Figures \ref{uniform_wt_image} and \ref{briggs_wt_image}.}
            \label{tgss_fig}
        \end{figure}

            To search for other counterparts to the diffuse emission detected with the GLEAM survey, we obtained an image for this part of the sky from the TIFR GMRT sky survey\footnote{\url{https://vo.astron.nl/tgssadr/q_fits/cutout/form}} \citep[TGSS, ][]{TGSS_paper} which is at a center frequency of $150$~MHz. This image did not show evidence for any supernova remnant (Fig.\,\ref{tgss_fig}).
            
            It is not surprising that the features were not visible in the TGSS (and ATCA) observations. The 150\,MHz TGSS image, shown in Fig.~\ref{tgss_fig}, has a comparable observing frequency to GLEAM \citep[][]{TGSS_paper} and similar RMS noise levels of a few mJy. However, in the case of both ATCA and TGSS, it is likely that they did not have sufficient u,v coverage to detect the diffuse structures found in GLEAM, which has ample sensitivity on short baseline spacings. Thus, with ATCA, TGSS and GLEAM, we are probing the region around \psr\ at drastically different spatial scales and frequency bands; it is clear based on TGSS and ATCA that the structure appears to be truly diffuse on scales approximately greater than a few arcminutes, thus lacks sharp filamentary structure.
            
            We did not find any emission associated with the GLEAM source in other wavelength bands using the {\sc ALADIN} \citep{aladin} sky atlas.
            The Chandra and XMM-Newton X-ray satellites do not have coverage for this portion of the sky. An analysis with the latest data from Fermi Large Area Telescope \citep[][]{LAT_desc}, accessed from the Fermi Science Support Center\footnote{The data access page can be found at the link: \url{https://fermi.gsfc.nasa.gov/ssc/data/}} \citep[][]{LAT_cat3}, also did not show evidence for a supernova remnant.
    
\section{Discussion}        \label{discuss}
    
    \subsection{Expectations}
        
        \begin{figure*}
            \centering
            \includegraphics[trim=5mm 1mm 20mm 15mm,clip,width = \textwidth]{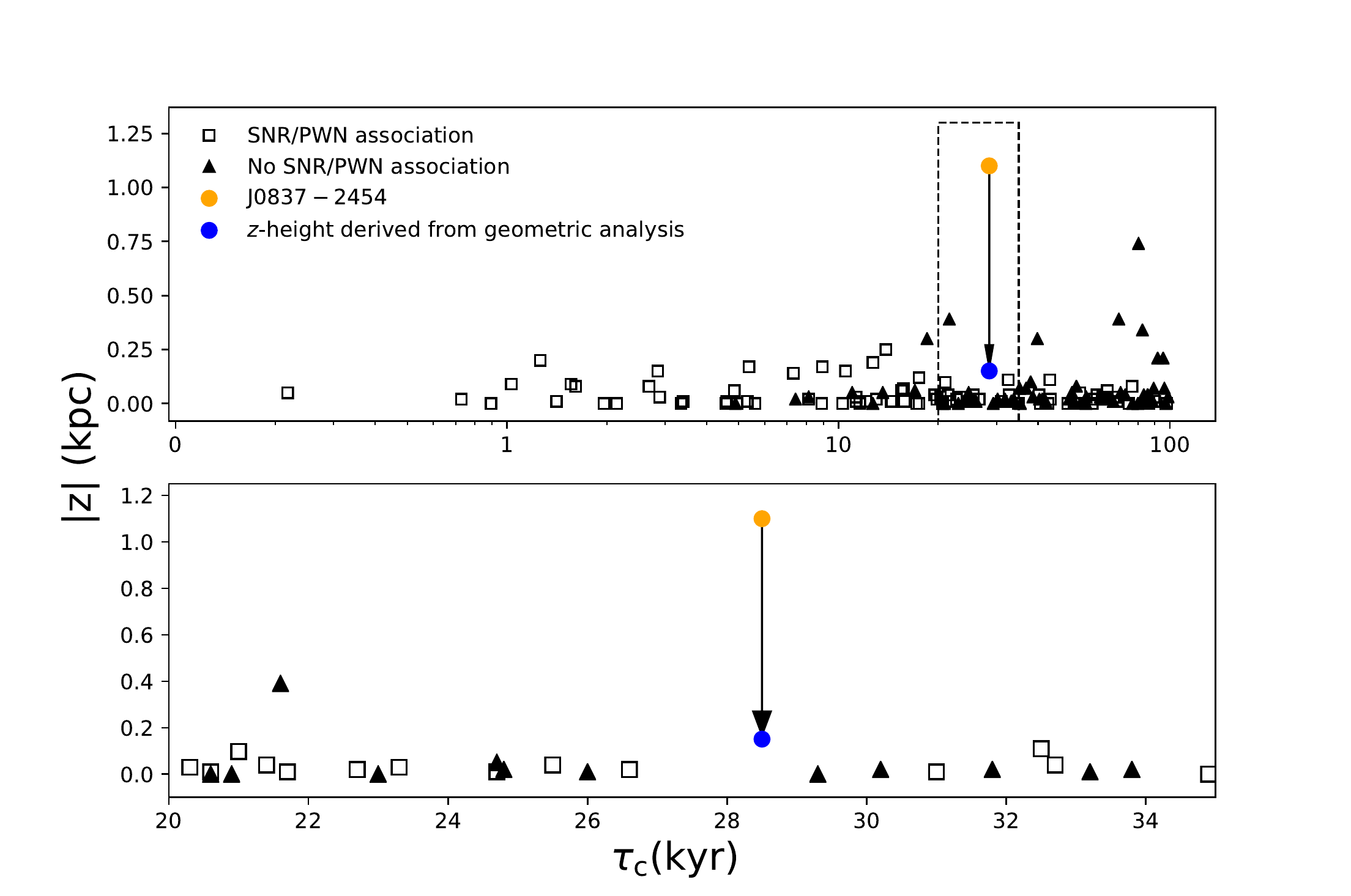}
            \caption{The absolute $z$-height of pulsars, inferred from the NE2001 model and their DMs, with $\tau_c\leq100$~kyr is plotted against their characteristic age in the top panel. The position of \psr\ based on the NE2001-derived $z$-height is shown with an orange dot. The position of \psr\ based on the geometric analysis in Sec.~\ref{new_z_ht} is shown with a blue dot. The bottom panel shows a zoom-in around the position of \psr, shown with the rectangle in the top panel. We can see that the NE2001-derived $z$-height of \psr\ is anomalously high compared to other pulsars of similar characteristic age. We can also see that many of the pulsars with similar age as \psr\ have an SNR/PWN associated with them.}
            \label{z_v_age}
        \end{figure*}
        
        Young pulsars like \psr\ are often associated with an SNR or a PWN. In Fig.~\ref{z_v_age}, we plot the Galactic $z$-height for all pulsars with a characteristic age $\tau_{\rm c} \leq 100$~kyr, highlighting the pulsars associated with a SNR/PWN. We can see that \psr\ has an anomalously high $z$-height as compared to other pulsars of similar age. Of the 74 pulsars with characteristic age $\tau_{\rm c} \leq 28.6$~kyr, 59 (i.e. 79.7\%) are associated with either an SNR or PWN.
        
        It is notoriously difficult to predict the properties of any potential SNR with any certainty. The PWN will typically be constrained to the region close to the pulsar, while the SNR will be more extended with a diameter ranging anywhere from few tens of arcseconds to a few degrees, depending on the expansion velocity, its stage of evolution, and the magnetic field and density of the interstellar medium. Even though the size and morphology of the SNR might be difficult to predict, assuming the characteristic age of \psr\ is a reasonable approximation of the age of the pulsar, we can expect the pulsar to be in the late Sedov expansion phase \citep[][]{sedov_expansion} or early radiative phase \citep{radiative_phase_snr}.
        
    \subsection{Is the GLEAM source an associated supernova remnant?}
        
        As discussed in Sec.~\ref{results}, we do not detect an SNR in our ATCA observations, however on a larger scale we detect extended emission in the lower-frequency GLEAM data. This emission is concentrated around \psr\ and shows morphology consistent with a diffuse, low-surface-brightness SNR.
        
        As described in Sec.~\ref{gleam_analysis}, we could not effectively assess the spectral index of this diffuse feature with the GLEAM data, thereby preventing us from making a conclusive statement about whether the diffuse source is the SNR. While \psr\ is also surrounded by H$\alpha$ emission (see Sec.~\ref{shassa_sec}), the difference in the morphology of the H$\alpha$ emission and the GLEAM diffuse emission suggests different origins for these two features. Thus, given the available data, we cannot confirm if the GLEAM diffuse emission is an SNR.
        
    \subsection{Limit on the Distance to the PSR/SNR Association} \label{new_z_ht}
        
        Despite the lack of a confirmed SNR around \psr, we can set limits on the distance to the pulsar using the observational evidence obtained from the SHASSA and GLEAM data and compare it to the one derived using the NE2001 model \citep{ne2001}. A distance discrepancy, if it exists, is not an unusual occurence in pulsar distance estimation \citep[see, for example,][]{wrong_dmdist_1, wrong_dmdist_2}. In particular, \citet[][]{wrong_dmdist_1} showed that the NE2001 model can over-estimate the distances to some pulsars, with a striking example of the DM-derived distance to PSR B0450--18 being a factor of $\sim$3 greater than the parallax distance. It is possible that the YMW model also suffers from similar over-estimation of distances to pulsars as the NE2001 model. If the NE2001 model overestimates the distance to \psr, that might explain why \psr\ is so unusual in its Galactic position relative to other young pulsars (see Figs.~\ref{fig:gcdist} and \ref{z_v_age}).
        
        \subsubsection{Using SHASSA H$\alpha$ emission} \label{shassa_dist}
            
            As described in Sec.~\ref{shassa_sec}, \psr\ is co-located with a diffuse H$\alpha$ emission complex. Some of this H$\alpha$ complex also lies along the line of sight to the pulsar and could be 
            contributing to the observed DM for the pulsar. This excess DM could lead to an overestimated distance, thereby overestimating the $z$-height of the pulsar. Since the NE2001 model does not list any over-densities or clumps in the free electron density in the direction of this pulsar \citep{ne2001}, it is not implicitly accounting for any excess DM that might be originating from the observed H$\alpha$ complex.
            
            From Fig.~\ref{shassa}, the amplitude of the H$\alpha$ emission from the pixel coincident with \psr's position is $\mathcal{R} = 470$~Rayleigh. Following the discussion in \citet[][]{frb_sparker} (their section 4.4), the SHASSA photon flux in Rayleighs can be converted to an emission measure, $\rm{EM} = \mathcal{R} \times 2.2 $~pc~cm$^{-6} = 1034$~pc~cm$^{-6}$. However, since we don't know the length of this H$\alpha$ complex along the line of sight, it is not possible to directly convert this EM to its corresponding free electron density or a dispersion measure contribution.
            
            Instead, we can place limits on the distance to \psr\ using a method similar to the one used in \citet[][]{BM_dist_limits}. Given a free electron density, $n_{\rm e}$, a filling factor, $f$, and the distance to the pulsar, $D_{\rm p}$, the dispersion measure can be calculated as, ${\rm DM} = n_{\rm e} \, f \, D_{\rm p}$. Similarly, the EM can be calculate as, ${\rm EM} = n_{\rm e}^2 \, f \, D_{\rm H\alpha}$, where $D_{\rm H\alpha}$ is the length of the H$\alpha$ complex.
            
            We can infer the approximate Galactic free electron density along the line of sight to the pulsar by dividing the DM estimated by the NE2001 model by the corresponding distance associated with this DM. This leads to a free electron density approximately an order of magnitude lower than the average mid-plane density of $\approx$0.1~cm$^{-3}$. Correspondingly, the Galactic estimate of the EM in this direction is negligible compared to that measered by SHASSA. Thus, we can assume that most of the observed DM and EM is a result of the H$\alpha$ complex along the line of sight.
            Assuming an approximately constant free electron density and filling factor in the this H$\alpha$ complex, the distance to the pulsar can be written as \citep{BM_dist_limits},
            \begin{equation}
                \displaystyle D_{\rm p} = \frac{DM}{\sqrt{EM}} \, \frac{\sqrt{D_{\rm H\alpha}}}{\sqrt{f}}
                \label{D_p_eq}
            \end{equation}
            Re-arranging the above equation and using the measured values of DM = 143~pc~cm$^{-3}$, EM = 1034~pc~cm$^{-6}$, and normalizing to a filling factor value of $f = 0.1$, we find that
            \begin{equation}
                \displaystyle \frac{D_{\rm p}}{D_{\rm H\alpha}} = 14 \, \left( \frac{0.1}{f} \right)^{1/2} \, (D_{\rm H\alpha})^{-1/2}.
            \end{equation}
            Assuming the H$\alpha$ complex fills the entire path along the line of sight to the pulsar, i.e.,
            \begin{equation}
                \displaystyle \frac{D_{\rm p}}{D_{\rm H\alpha}} \sim 1,
            \end{equation}
            gives us an estimate of the distance to the pulsar,
            \begin{equation}
                \displaystyle D_{\rm p} \approx 0.2 \, \left( \frac{0.1}{f} \right)^{1/2} {\rm kpc}.
            \end{equation}
            This distance is significantly smaller than that predicted by NE2001 model ($D_{\rm cl} = 6.3$~kpc). This distance corresponds to a $z$-height of $\approx$34~pc, which is consistent with those observed for other young pulsars like \psr.
            
            However, if the H$\alpha$ complex does not fill the entire distance out to the pulsar, as we assumed in this analysis, then given the small Galactic free electron density along this line of sight, the pulsar could be much further away than the distance calculated above. This effect would also dominate the changes in the distance estimate arising from differences in the assumed filling factor for the H$\alpha$ complex.
            
        \subsubsection{Using GLEAM diffuse emission} \label{gleam_dist}
        
        If we assume that the diffuse emission observed in the GLEAM data is indeed the SNR, then we can infer a limit on the distance to \psr\ by using the average velocity of expansion of the SNR shell in different stages of evolution. From Fig.~\ref{fig:gleam}, an angular radius that visually encloses most of the SNR emission can be approximated as $\theta \sim 0.7^{\circ}$. Using the small angle approximation, the spatial radius of the SNR will be $R = D \times \theta$, where $D$ is the distance to the pulsar-SNR system. We also know that the approximate age of the pulsar-SNR system is $\tau_{\rm c} = 28.6$~kyr.
        Next, we assume an initial ejecta velocity of $v_{\rm e}=5000$~km/s for our SNR in the ejecta-driven phase lasting $t_{\rm e}\simeq 1$~kyr and a typical $v_{\rm s}=200$~km/s during the $27.6$~kyr since then, which the SNR has spent in the Sedov expansion phase \citep[][]{handbook_sne}. Combining the two velocities, we get a velocity estimate that is weighted by the duration of time spent in each evolutionary phase of
        \begin{equation}
            \bar v = \frac{v_{\rm e}(t_{\rm e})}{\tau_{\rm c}} +  \frac{v_{\rm s}(\tau_{\rm c}-t_{\rm e})}{\tau_{\rm c}} \approx 370 \, {\rm km/s}
            \label{vel_eq}
        \end{equation}
        
        Using the average velocity and time for which the SNR has been expanding, we can determine the distance to the SNR:
        \begin{equation}
        \begin{split}
            D = \frac{\bar v \tau_{\rm c}}{\theta}
        \end{split}
        \label{distance_eq}
        \end{equation}
        Assuming the values previously noted, we determine a distance of 
        \begin{equation}
            \displaystyle D \approx 0.9 \, \left( \frac{\bar v}{370~{\rm km/s}} \right) \left( \frac{\tau_{\rm c}}{28.6~{\rm kyr}} \right) \left( \frac{0.7^{\circ}}{\theta} \right) \, {\rm kpc}
        \end{equation}
        This distance also differs significantly from the predictions made using the NE2001 model ($D_{\rm cl} = 6.3$~kpc).
        
        The above analysis is based on simple constant weighted velocity and geometric arguments and is independent of the Galactic free electron density in that direction in the Milky Way. As such, it provides an independent estimate of the distance to the pulsar-SNR system. As we discuss below, for the distance derived from this analysis to agree with the distance derived from the NE2001 model, the SNR ejecta would need to have a much higher average velocity, or the pulsar and SNR would have to be much older than the derived characteristic age. Another way in which this distance discrepancy could be resolved is if the diffuse GLEAM source is not in fact the SNR.
        
        Fixing the distance in Eq.~\ref{distance_eq} to be equal to that predicted by the NE2001 model, i.e. $D = D_{\rm cl} = 6.3$~kpc and assuming the characteristic age is a reasonable approximation, that implies that the average velocity of the SN ejecta over the past 28.5~kyrs would have to be approximately 2600\,km\,s$^{-1}$. This average velocity implies, using Eq.~\ref{vel_eq} and the same assumption of $v_{\rm e}=5000$~km/s for our SNR in the ejecta-driven phase lasting $t_{\rm e}\simeq 1$~kyr, that the ejecta would need to be moving at a velocity of $\sim$2500\,km\,s$^{-1}$ in the post-ejecta-driven phase. This velocity seems uncharacteristically high for ejecta that are out of their ejecta-driven phase \citep{handbook_sne} and thus is unlikely to be representative of the true velocity of the ejecta. Based on these arguments, a distance much smaller than that predicted by the NE2001 model is preferred for the pulsar-SNR system.

        Similarly, to be consistent with the distance estimate from the NE2001 model, the pulsar-SNR system would have to be a factor of $\sim$7 older than the characteristic age. It is not unusual for a pulsar to have a true age to be significantly different from the characteristic age. For example, PSRs B1951+32 and B0538+2817 have true ages that are a factor of 0.5 and 0.05 smaller than their characteristic ages respectively \citep[][respectively]{B1951_true_age, B0538_true_age}. Another example of such a discrepancy is PSR J1801--2451, where the pulsar's proper motion suggests that the true age of this pulsar could be significantly higher than its characteristic age \citep{J1801_true_age}. Two sources of uncertainty in the calculation in the characteristic age (Eq.~\ref{char_age}) are the braking index, $n$, and the period derivative, $\dot{P}$. In our calculation, we assumed $n = 3$, though it is possible that the braking index for this pulsar might be significantly different \citep{noutsos_braking_index}. To get an age which is a factor of $\sim$7 greater than that we calculate in Eq.~\ref{approx_char_age}, the braking index would have to be $n \approx 1.3$ which seems unlikely, given that only the Vela pulsar \citep[$n \approx 1.4$,][]{vela_braking_index} and PSR J1734--3333 \citep[$n \approx 1$,][]{1734_braking_index} have been measured to have such low breaking indices. Finally, we might be underestimating the characteristic age if we were overestimating the period derivative, $\dot{P}$, of the pulsar. This is possible given that the period derivative can be contaminated by the kinematic motion of the pulsar, in what is known as the ``Shklovskii effect" \citep{shklovskii_effect}. The contribution from the kinematic terms is usually small for pulsars like \psr, but given its unique location in the Galaxy, these effects might be significant enough to resolve the distance discrepancy with the NE2001 model described above. If the pulsar indeed turns out to be significantly older, there are twelve pulsars with characteristic ages greater than 100~kyr which have a supernova remnant associated with them, so it might still be possible to detect emission from this SNR.
        
        Assuming that a distance of $D = 0.9$~kpc is an accurate estimate of the true distance to \psr,\ we can calculate the $z$-height of the pulsar above the Galactic plane to be $z_{\rm new} = D \times {\rm sin} (b)$, 
        where $b$ is the latitude of \psr.\ Doing the calculation, we get a new $z$-height for \psr\ of $z_{\rm new} \approx 151$~pc. This new $z$-height is much more reasonable than the $z$-height predicted by the NE2001 model shown in Fig.~\ref{z_v_age}. This new $z$-height places it among other known pulsars of similar ages, as shown in Fig.~\ref{z_v_age}. 
        
    \subsection{What if the NE2001 position is accurate?}
    
        It is likewise possible that the diffuse GLEAM object is not a supernova remnant, and/or not associated with the pulsar's origin. As noted in the previous section, if the pulsar is indeed as distant as the DM models imply, any associated SNR would be expected to be at a location and size detectable by ATCA. In this scenario, we obviously did not detect the SNR, so we discuss here a few reasons why one might not have been detected, as well as the implications of a pulsar detected this far from the Galactic plane.
        
        \subsubsection{Implication of position of \psr}
        
        If accurate, the NE2001-derived position of \psr\ makes it unique among the known population of young pulsars in the Milky Way (see Fig.~\ref{z_v_age}). Given the large period and period-derivative of the pulsar, it is not a recycled pulsar and was likely born from a core-collapse supernova from an O/B-type star. These massive stars are located in the disk of the Galaxy with a scale height of 34~pc \citep{OB_star_spatial_dist_1}.
        
        We can estimate the $z$-velocity, $V_{\rm z}$, \citep{Maura_highb_psr} of the pulsar given its kinematic age, $\tau_{\rm k}$, and the z-height at which it was born, $z_{0}$,
        \begin{equation}
            \displaystyle V_{\rm z} = \frac{z - z_0}{\tau_{\rm k}},
            \label{z_vel_eq_1}
        \end{equation}
        where $z$ is the current $z$-height of the pulsar.
        Assuming the pulsar was born in the plane of the Galaxy, i.e. $z_0 = 0$, the true $z$-height of the pulsar is given by the NE2001-derived distance, $z = z_{\rm cl} = 1.1$~kpc (Table~\ref{table:timingsolution}), and that the kinematic age is equal to the characteristic age, $\tau_{\rm k} = \tau_{\rm c} = 28.6$~kyr, the $z$-velocity of the pulsar is,
        \begin{equation}
            \displaystyle V_{\rm z} \approx 3.8 \times 10^4 \left( \frac{28.6 \ {\rm kyr}}{\tau_{\rm k}} \right) \left( 1 - \frac{z_0}{1.1 \ {\rm kpc}} \right) \ {\rm km \ s^{-1}}.
            \label{z_vel_eq_2}
        \end{equation}
        This velocity is more than an order of magnitude greater than the tail of the pulsar velocity distribution \citep{og_vel_dist, psrpi_vel_dist}.
        
        Given the large velocity derived in Eq.~\ref{z_vel_eq_2}, it is unlikely that the pulsar was born in the plane of the Galaxy and migrated to its current observed position. However, majority of O/B stars are found to be in binaries \citep[][]{OB_in_binaries}. Most of these binaries are disrupted when one of the members undergoes a supernova explosion \citep{runaway_star_numerical_sim}. Assuming \psr\ was born from the surviving massive star in such a disrupted binary, the velocity imparted to the star during the disruption event could have allowed the progenitor O/B star to travel further away from the plane of the Galaxy before itself undergoing a supernova remnant to form \psr. If the progenitor of the pulsar was indeed such a runaway star, then in the most optimistic case, it would have a peculiar velocity of $\sim$60~km~s$^{-1}$ \citep{runaway_star_numerical_sim}, which when combined with an average lifetime of 30~Myr for O/B stars, would allow the progenitor to travel $\approx$2~kpc before collapsing to a neutron star. Thus a runaway star hypothesis for the progenitor of \psr\ would allow it to have a much more reasonable $z$-velocity by having the pulsar born much closer to its current observed position.
        
        Another way to reduce the $z$-velocity would be if the kinematic (or true) age of the pulsar were significantly larger than the characteristic age of the pulsar, which was already explored in Sec.~\ref{gleam_dist}. In fact, a combination of a larger kinematic age and the progenitor of the pulsar being the second star in a disrupted binary is also highly possible. This would also relax the condition that the progenitor be a runaway star and allow it to be a ``walkaway'' star, which are more common than runaway stars and have peculiar velocities  $\leq$30~km~s$^{-1}$ \citep{runaway_star_numerical_sim}. Thus, if the NE2001-derived distance is accurate, this would make this the first pulsar that is known to be born from either a runaway or walkaway progenitor star.
        
        These hypotheses can be verified through a measurement of the velocity of \psr. Because of the large timing residuals, we cannot measure the velocity of the pulsar through timing. Since young pulsars tend to exhibit lots of red noise in their timing, we will need a significant increase in the timing baseline to increase the probability of measuring a velocity through timing. The small flux density of the pulsar also inhibits proper motion and parallax measurements through VLBA, which requires pulsars with a flux density higher than 1.6~mJy \citep{Deller_vlbi}. Similarly, the low flux density would also make it difficult to observe scintillations of the pulsar emission and derive a velocity through interstellar scintillation.
        
        \subsubsection{Lack of SNR emission}
        
        Based on the arguments presented above, it is likely that the pulsar was born close to its current NE2001-predicted position. If the characteristic age of the pulsar is a good estimate of the true age of the pulsar, we would have expected a SNR in its late Sedov \citep{sedov_expansion} or early radiative evolution phase \citep{radiative_phase_snr}.
        However, in these regions of the Galaxy, it is possible that a low ambient circumstellar density at the position of \psr\ could contribute to the formation of a diffuse, low-surface-brightness SNR. Radio emission from SNRs arises from synchrotron emission; the relevant electrons are accelerated by the compression of the ambient magnetic field in the shock front, and the matter expelled from the supernova shocks the circumstellar medium \citep[diffusive shock acceleration,][]{particle_accel_snr_1, particle_accel_snr_2}. If the density of the circumstellar medium is inherently low, it will inhibit the formation of a shock-front; this would serve to both reduce the synchrotron emission from the SNR and to lead to a more diffuse observed structure due to a longer phase of rapid expansion. 
        
        As synchrotron radiation from SNRs also requires the presence of an ambient magnetic field, an insufficiently strong field will contribute to a weak synchrotron shock front. During a supernova, the ambient field is amplified to strengths sufficient for synchrotron emission \citep[see][for a review of different amplification processes]{mag_field_amp_1, mag_field_amp_2}. \citet[][]{mag_field_amp_2} analyzed archival supernovae to show that the post-amplified strength of magnetic fields in these SNRs ranged from $25$~to~$1000 \ \upmu$G, or about 2.5--100 times higher than the diffuse Galactic magnetic field. \citet[][]{gal_mag_field} created a map of the galactic magnetic field using the NE2001 thermal electron density model, incorporating the findings of \citet[][]{gaensler_wim_structure}. Based on their analysis \citep[see Eq.~1][]{gal_mag_field}, the ambient Galactic magnetic field strength around \psr\ will be on the order of $\sim 2 \, \upmu$G corresponding to the distance prediction made by the NE2001 model. This magnetic field is approximately five times lower than the ambient magnetic field in the disc of the Galaxy, and would require a highly efficient process to amplify it to the field strengths observed in other supernova remnants. A combination of low circumstellar density and low ambient magnetic field would thus work to lower the expected flux density of any SNR, thus making it much harder to detect.
        
\section{Conclusions}       \label{conclusion}
    
    We present the timing solution for \psr\ using the Parkes radio telescope. The timing solution reveals that this pulsar exhibits red noise, and has a relatively young characteristic age of $\tau_{\rm c} \approx 28.6$~kyr. Given the young age of the pulsar, we search for a supernova remnant around \psr\ with the ATCA telescope. We detect the pulsar in our ATCA data and discovered low surface brightness diffuse emission surrounding the pulsar in archival GLEAM data. The low surface brightness diffuse emission, which if we assume to be the SNR, implies a low local free electron density in the pulsar environment, which is consistent with the prediction from the NE2001 model. Based on a geometrical analysis of a putative SNR identified at low frequencies, we derive a distance of $\sim$0.9~kpc to the pulsar-SNR system.
    We also detect co-located H$\alpha$ emission from the SHASSA survey at the position of the pulsar. The morphology of this H$\alpha$ emission is inconsistent with the morphology of the GLEAM diffuse emission. Assuming that the excess DM comes from this H$\alpha$ complex, we estimate the distance to the pulsar to be $\sim$0.2~kpc.
    Both of these distance estimates are significantly smaller than the predictions based on the NE2001 and YMW models, highlighting the uncertainty in estimating distances to pulsars using dispersion measures. The new distance estimates also result in $z$-heights for the pulsar which are consistent with other pulsars of similar age.
    
    The discrepancies in the distance estimates based on the electron density models and the geometric approach point to an important caveat relating to the identification of Fast Radio Bursts (FRBs). The fundamental basis of ``extragalactic'' FRB identification relies on determining the Galactic contribution to the total DM observed for FRBs. The significant error in the leading electron densities toward the anti-center that are implied by our observations (factors of 10--20) indicate that care must be taken in the interpretation of transient discoveries (repeating and non-repeating) as Galactic or extragalactic towards this region of the sky \citep{keane_classn_frb}.
    
    However, if the NE2001-derived distance is accurate, then that, coupled with the young age of the pulsar, would imply that the pulsar was likely born close to its current observed position. This would imply that the progenitor O/B star would have to be a runaway star, making this the first pulsar known to be born from a runaway progenitor star.

\acknowledgements

This research has made use of the NASA/IPAC Extragalactic Database (NED) which is operated by the Jet Propulsion Laboratory, California Institute of Technology, under contract with the National Aeronautics and Space Administration. This research has also made use of the SIMBAD database, operated at CDS Strasbourg, France. This research has made use of "Aladin sky atlas" developed at CDS, Strasbourg Observatory, France

NP, SBS and HB are members of the NANOGrav Physics Frontiers Center (NSF PHY-1430284). NP is also supported by NSF AAG-1517003. EP acknowledges funding from an NWO Veni Fellowship. NHW is supported by an Australian Research Council Future Fellowship (project number FT190100231) funded by the Australian Government.

The Australia Telescope Compact Array and Parkes radio telescope are part of the Australia Telescope National Facility which is funded by the Australian Government for operation as a National Facility managed by CSIRO. This project was in part funded by NSF EPSCoR award number 1458952. The Southern H-Alpha Sky Survey Atlas (SHASSA), which is supported by the National Science Foundation 

\software{dspsr \citep[][]{dspsr}, psrchive \citep[][]{psrchive}, CASA  \citep[][]{casa}, Aladin \citep[][]{aladin_1, aladin_2}}

\appendix

\section{Point sources detected in ATCA field-of-view}
    
    In Table~\ref{sources}, we list the point sources that were detected in the ATCA field of view along with their association to known point sources in the NVSS and SIMBAD catalog.
    
    \begin{table*}[htb]
        \centering
        \begin{tabular}{c c c c c}
            \toprule
            Sr. No. & Ident. & RA & DEC & Integrated Flux \\
            & & (hh:mm:ss.sss) & (dd:mm:ss.ss) & (mJy) \\
            \midrule
            1 & NVSS 083817-251133 &    08:38:17.038 $\pm$ 0.022 s & -25:11:40.028 $\pm$ 0.142 s & 37.60 $\pm$ 2.300 \\
            2. & NVSS 083732-244611 &    08:37:32.049 $\pm$ 0.014 s & -24:46:16.948 $\pm$ 0.087 s & 16.57 $\pm$ 0.630 \\
            3. & NVSS 083839-245106 &    08:38:39.814 $\pm$ 0.012 s & -24:51:09.144 $\pm$ 0.076 s & 14.77 $\pm$ 0.460 \\
            4. & NVSS 083814-245127 &    08:38:14.333 $\pm$ 0.008 s & -24:51:28.966 $\pm$ 0.057 s & 8.960 $\pm$ 0.200 \\
            5. & NVSS 083801-245739 &    08:38:01.349 $\pm$ 0.007 s & -24:57:42.372 $\pm$ 0.043 s & 5.049 $\pm$ 0.093 \\
            6. & NVSS 083746-245530 &    08:37:46.651 $\pm$ 0.012 s & -24:55:33.085 $\pm$ 0.076 s & 4.190 $\pm$ 0.130 \\
            7. & NVSS 083853-244609 &    08:38:54.054 $\pm$ 0.050 s & -24:46:17.975 $\pm$ 0.210 s & 3.170 $\pm$ 0.290 \\
            8. & NVSS 083751-244811 &    08:37:52.272 $\pm$ 0.025 s & -24:48:13.889 $\pm$ 0.132 s & 2.840 $\pm$ 0.170 \\
            9. & NVSS 083654-244656 &    08:36:54.079 $\pm$ 0.082 s & -24:46:53.655 $\pm$ 0.418 s & 2.700 $\pm$ 0.640 \\
            10. & NVSS 083648-245309 &    08:36:48.320 $\pm$ 0.021 s & -24:53:16.525 $\pm$ 0.247 s & 0.965 $\pm$ 0.140 \\
            11. & -- &    08:38:53.226 $\pm$ 0.023 s & -24:58:45.749 $\pm$ 0.254 s & 0.894 $\pm$ 0.101 \\
            12. & NVSS 083713-245111 &    08:37:13.842 $\pm$ 0.043 s & -24:51:12.270 $\pm$ 0.326 s & 0.774 $\pm$ 0.114 \\
            13. & -- &    08:37:51.795 $\pm$ 0.051 s & -24:58:51.020 $\pm$ 0.327 s & 0.494 $\pm$ 0.079 \\
            14. & -- &    08:37:41.857 $\pm$ 0.039 s & -24:51:36.157 $\pm$ 0.236 s & 0.444 $\pm$ 0.049 \\
            15. & -- &    08:38:00.101 $\pm$ 0.057 s & -24:51:54.951 $\pm$ 0.364 s & 0.431 $\pm$ 0.069 \\
            16. & -- &    08:37:49.562 $\pm$ 0.041 s & -24:50:32.491 $\pm$ 0.243 s & 0.343 $\pm$ 0.048 \\
            17. & -- &    08:37:48.019 $\pm$ 0.074 s & -24:56:57.376 $\pm$ 0.451 s & 0.311 $\pm$ 0.059 \\
            18. & -- &    08:37:23.562 $\pm$ 0.008 s & -24:48:04.798 $\pm$ 0.161 s & 0.305 $\pm$ 0.034 \\
            & & & & \\
            19. & \psr\     &    08:37:57.723 $\pm$ 0.086 s & -24:54:27.599 $\pm$ 0.571 s & 0.174 $\pm$ 0.048 \\
            & & & & \\ 
            \bottomrule
        \end{tabular}
        \caption{The point sources detected in the uniform weighted image (Fig.~\ref{uniform_wt_image}) along with their association to known point sources in the NVSS and SIMBAD catalogs. In the case where the source had been identified in both catalogs, we list the identification associated with the NVSS catalog.}
        \label{sources}
    \end{table*}

\bibliographystyle{aasjournal}
\bibliography{bibliography}

\end{document}